\documentclass[twocolumn]{aastex63}
\usepackage{graphics,epsf}
\usepackage{amsmath}                
\usepackage{amsfonts}               
\usepackage{amssymb}                
\usepackage{epsfig}                 
\usepackage{appendix}
\usepackage{graphicx}
\usepackage{float}
\usepackage{color}
\usepackage{multirow}
\usepackage{colortbl}
\usepackage[para,online,flushleft]{threeparttable}

\newcommand{\kms}{{~\rm km\; s^{-1}}}

\newcommand{\m}{{~\rm m}}
\newcommand{\km}{{~\rm km}}
\newcommand{\s}{{~\rm s}}
\newcommand{\h}{{~\rm h}}

\newcommand{\g}{{~\rm g}}
\newcommand{\G}{{~\rm G}}
\newcommand{\K}{{~\rm K}}

\newcommand{\Myr}{{~\rm Myr}}

\newcommand{\kpc}{{~\rm kpc}}

\usepackage{lineno}

\begin{document}

\title{Comment on "A snapshot of the oldest AGN feedback phases"}

\author[0000-0003-0375-8987]{Noam Soker}
\affiliation{Department of Physics, Technion, Haifa, 3200003, Israel; \\ soker@physics.technion.ac.il, shlomi.hillel@gmail.com}

\author{Shlomi Hillel}
\affiliation{Department of Physics, Technion, Haifa, 3200003, Israel; \\ soker@physics.technion.ac.il, shlomi.hillel@gmail.com}

\begin{abstract}
We dispute a very recent claim, which is based on new LOFAR radio observations, that mixing does not heat the intragroup medium of the galaxy group Nest200047. We argue to the contrary, namely, that the main heating process is mixing, by showing that the radio morphology of filaments in this galaxy group was qualitatively reproduced by a past three-dimensional hydrodynamical simulation that also showed that the main heating process is by mixing of hot gas from the jet-inflated bubbles with the intracluster (or intragroup) medium. 
\end{abstract}

\section{INTRODUCTION}
\label{sec:intro}

There is an ongoing dispute on the main heating process of the intracluster medium (ICM) in clusters of galaxies and of the intragroup medium in group of galaxies (for a recent relevant review see \citealt{Soker2016}). 

In one group of heating processes the jets that the central active galactic nucleus (AGN) launches and the bubbles that the jets inflate do work on the ICM by driving shocks (e.g., \citealt{Randalletal2015, Guoetal2018}),  by exciting sound waves (e.g., \citealt{Fabianetal2017, TangChurazov2018}), by powering turbulence (e.g.,    \citealt{DeYoung2010, Gasparietal2014, Zhuravlevaetal2017}), and/or by uplifting gas from inner regions (e.g., \citealt{GendronMarsolaisetal2017}). 
However, several studies point to some problems with these processes (for a recent discussion see \citealt{Soker2019CR, HillelSoker2020}). 

In the second group of heating processes there is a heat transport from the hot jet-inflated bubbles to the ICM, including streaming of cosmic rays (e.g. \citealt{FujitaOhira2013, Pfrommer2013, JiangOh2018, Ruszkowskietal2018}) and heating by mixing. In the heating by mixing process many vortices, which the jets and the bubbles they inflate form,  mix hot bubble gas into the ICM (e.g., \citealt{BruggenKaiser2002,  Bruggenetal2009, GilkisSoker2012, HillelSoker2014, YangReynolds2016b}). The heating by mixing is more efficient than the cosmic ray streaming \citep{Soker2019CR}. 

In previous studies we (e.g., \citealt{HillelSoker2017a, HillelSoker2018, HillelSoker2020} for the latest papers) argued that the heating by mixing is the main heating process of the ICM in cooling flow clusters and groups. All other heating processes that we listed above do take place, like excitation of sound waves, shocks, and ICM turbulence, but our view is that they are sub-dominant relative to heating by mixing.
 
In a recent paper \cite{Brienzaetal2021} state that ``However, this lack of mixing by no means reduces the efficiency of the AGN feedback,
since the energy exchange between the bubbles and the intra-group medium can proceed without a thermal coupling of these phases.''.
We disagree with this conclusion. 

We show below that our previous three-dimensional hydrodynamical simulation where we found that heating by mixing is the most efficient heating process qualitatively resembles the general radio morphology they observe in the galaxy group Nest200047. 
We therefore argue to the contrary of the above claim by \cite{Brienzaetal2021}. Namely, we argue that their results actually support the heating by mixing process. 

\section{The numerical simulations}
\label{sec:Simulations}

We present results from our earlier three-dimensional hydrodynamical simulation \citep{HillelSoker2016}, which we also analysed in recent papers where we give more details (\citealt{HillelSoker2017a, HillelSoker2018, HillelSoker2020}). We performed these simulations with the numerical code {\sc pluto} \citep{Mignone2007}. 
The essential parameters for this short comment are as follows. To save numerical resources we simulated only the octant, $x>0$, $y>0$ and $z>0$. The highest resolution of this adaptive mesh refinement grid is $\approx 0.1 \kpc$. This implies that we cannot resolve vortices that are smaller than $\approx 1 \kpc$ at best. One should bear this limited resolution in mind. 

We launched slow-wide jets (e.g., \citealt{Aravetal2013} for observational support) with a half-opening angle of $\theta_{\rm j} = 70^\circ$ along the $z$ axis with an initial velocity of $v_{\rm j} = 8200 \kms$. The jets outlet is a circle $\sqrt{x^2 + y^2} \leq 3 \kpc$  at the plane $z = 0$. 
The jets (we simulate one jet of the two opposite jets) are periodic with an active phase of $10 \Myr$ followed by a quiescence (no jets) period of $10 \Myr$. The jet-active phases are in the time intervals of 
\begin{equation}
20(n-1) \le t_n^{\rm jet} \le 10 (2n-1), \qquad n=1,2,3 \dots .
\label{eq:jet}
\end{equation}

\begin{figure*}
\begin{center}
\includegraphics[trim=9cm 2.5cm 9cm 4.00cm,scale=0.84]{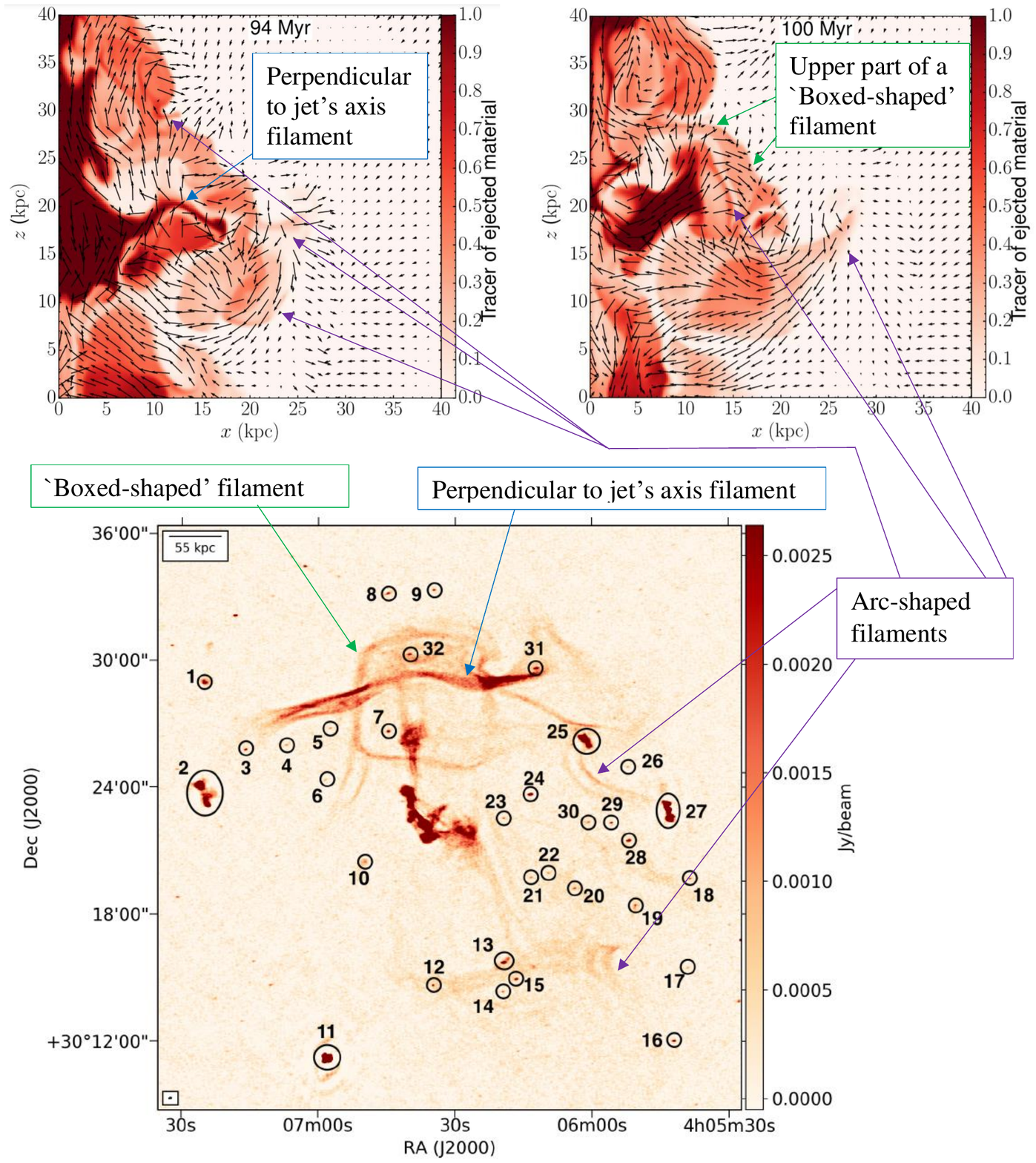} \\
\caption{Upper panels. Results from the three-dimensional hydrodynamical simulation by \cite{HillelSoker2017b} showing the meridional plane $y = 0$ at two times as indicated in the panels. The white-red colors indicate the value of the tracer of the jets $\xi_j$ according to the color coding bar on the right. A value of $\xi_j=1$ implies pure jet's gas while $\xi_j=0$ indicates only intracluster gas. The arrows indicate the velocity direction and magnitude emphasising the low (sub-sonic) velocity structure.
Namely, all velocities of $v>400 \km \s^{-1}$ are marked with the same maximum length of the arrows.   
Lower panel (from \citealt{Brienzaetal2021}): A LOFAR radio image at 144~MHz of the galaxy group Nest200047. Circles mark compact sources in the field.  }
\label{fig:Schematic}
\end{center}
\end{figure*}

This is the same simulation that we presented in \cite{HillelSoker2017b} where we give more details.
We take the two upper panels of Fig. \ref{fig:Schematic} from figure 3 of that paper. These two panels present the tracer of the jets and the velocity maps.  
The tracer is a non-physical variable that is frozen-in to the flow, and therefore marks the origin of the gas. Here we take the tracer to follow that original gas of the jet. At the origin of the jets $\xi_j = 1$. The rest of the gas in the grid (that does not come from the jets), i.e., the ICM (or intragroup gas) starts at $t=0$ with $\xi_j (0) = 0$. At later times $\xi_j (t,x,y,z)$ is the fraction of the gas that started in the jets and now is inside the cell at $(x,y,z)$. 

We take the tracer to indicate the location of radio emission, as the radio emission results from high energy electrons that originate in the shocked jets' gas. The tracer location marks the location of the radio emission, but we do not calculate the intensity here.  Therefore, we can only follow the morphology, but neither the radio intensity nor the spectral index that indicates the age of the high-energy electrons. We encourage new high-resolution simulations that include the calculation of the radio emission and magnetic fields.   
    
\section{Morphology}
\label{sec:Morphology}

When comparing the observed radio morphology from \cite{Brienzaetal2021} that we present in the lower panel of Fig. \ref{fig:Schematic}) with the numerical simulation (two upper panels) we should bear in mind the following. 
\begin{enumerate}
  \item The limited resolution of the numerical simulation. 
  \item That the panels from the simulation present only the tracer of the jets in the meridional plane and not the radio intensity. 
  \item That the morphology changes with time and location. We see this in the two upper panels that are close in time but still presents different morphologies. This is more pronounced in the observations that present largely two different morphologies on the north and south of the image. Namely, the `Boxed-shaped' filament and the perpendicular filament appear only on the northern side, and the faint filaments on the north have different distribution than what the faint filaments on the south have. 
  \item The morphology due to the mixing of hot jet-inflated bubble gas with the ICM (or intragroup medium) is sensitive to the manner of jets' activity. The simulation contains only one such activity cycle (equation \ref{eq:jet}). The jets' activity cycles of the galaxy group Nest200047 might have been completely different.   
  \item As \cite{Brienzaetal2021} mention, magnetic fields might play a significant role in shaping the filaments. The numerical simulation does not include magnetic fields.  
\end{enumerate}

Overall, we cannot make neither a quantitative comparison nor a one-to-one qualitative comparison between our old simulation and the new observations. However, we can clearly state that the simulation can produce some features that are observed. We mark these features on Fig. \ref{fig:Schematic}. These include curved (`arc-shaped') filaments, narrow features perpendicular to the jets' axis, and large filaments that start perpendicular to the jets' axis and then sharply bend by about $90^\circ$. When closed on the other side of the symmetry axis (that we do not simulate) this filament forms the upper part of a box. 

Our main claim from this comparison is that the mixing of shocked-jets' material, i.e., the hot gas of the jet-inflated bubbles, with the intragroup medium can form the general radio morphology of the galaxy group Nest200047.
In the past (e.g., \citealt{HillelSoker2016}) we used the same simulation to argue that this mixing is the main heating process of the ICM or intragroup medium. Namely, the  heating by mixing process is the dominant heating process (although other processes also take place).

\section{Summary}
\label{sec:Summary}

We presented arguments to the contrary of the claim of \cite{Brienzaetal2021} that mixing does not heat the intragroup medium of the galaxy group Nest200047. 
By comparing the observed radio morphology of the galaxy group Nest200047 from \cite{Brienzaetal2021} with a three-dimensional hydrodynamical simulation from \cite{HillelSoker2016} we argued that our simulation of jet-inflated bubbles in the ICM (and intragroup medium) can reproduce the main morphological features of the radio observations. This type of jet-inflated bubbles by wide (or precessing) jets show also that heating by mixing is the main heating process of the ICM (e.g., \citealt{GilkisSoker2012, HillelSoker2016, Sokeretal2016}). 
We therefore conclude, contrary to the claim of \cite{Brienzaetal2021}, that the new radio observations of the galaxy group Nest200047 support the claim that heating by mixing is the main heating process of the intragroup medium.
 
 Moreover, our long simulations for $t > 100 \Myr$ in two-dimensions \citep{HillelSoker2014} and in three-dimensions \citep{HillelSoker2016} show that the bubbles still exist, despite the vigorous mixing of some fraction of the hot bubble gas with the ICM. The key process is to inflate the bubbles self-consistently by jets \citep{SternbergSoker2008}. The bubbles survive despite that we do not include magnetic fields. This is also along the new finding by \cite{Brienzaetal2021}.

\label{lastpage}
\end{document}